\documentclass[10pt,letterpaper]{article}
\usepackage{opex3}
\usepackage{textcomp}
\usepackage{subfigure}
\usepackage{cite}
\usepackage{SIunits} 
\begin{document}

\title{Feasibility of underwater free space quantum key distribution}
\author{Peng Shi, Shi-Cheng Zhao, Wen-Dong Li and Yong-Jian Gu$^{*}$}
\address{Department of Physics, Ocean University of China, Qingdao 266100,
China}
\email{yjgu@ouc.edu.cn}

\begin{abstract*}
We investigate the optical absorption and scattering properties of
underwater media pertinent to our underwater free space quantum key distribution (QKD)
channel model. With the vector radiative transfer theory and Monte Carlo
method, we obtain the attenuation of photons, the fidelity of the scattered photons, the quantum bit
error rate and the sifted key generation rate of underwater quantum
communication. It can be observed from our
simulations that the most secure single photon underwater free space QKD is feasible
in the clearest ocean water.
\end{abstract*}




\section{Introduction}

Driven by the communication requirements of underwater sensor networks,
submarines and all kinds of underwater vehicles, underwater wireless optical
communication has been developing rapidly in recent years \cite%
{MTivey,IVasilescu,FHanson,NFarr,JJPuschell}. The new research shows that by
using the current technology underwater wireless optical
communication can transmit $350m$ in the clearest
seawater with the bit rate of $10Mbps$ \cite{MLanzagorta}. At the same time,
strong research effort has been devoted to the study of how quantum effects
may be employed to manipulate and transmit information, that is called
quantum information processing \cite{MANielsen}. Quantum key distribution
(QKD) is an important branch of quantum information processing, in
particular, is on its way from research laboratories into the real world.
Consequently, QKD could be used to provide security for underwater wireless
optical communication. Recent years, QKD based on photons makes great
progress both in theoretical and experimental researches \cite{NGisin}. The
BB84 protocol is the first protocol to describe how to establish a secret
key with polarization encoded photons as qubits \cite{VScarani09,CHBennett}. QKD has been proved absolutely safe by some fundamental
principles of quantum physics \cite{CHBennett,AKEkert,SJWiesner}, and the
maximum distance of QKD using single photons in free space has been reached $%
144km$ \cite{SMTobias}. It provides an important basis for the research of
optical underwater free space QKD. 

Just like quantum communication in free atmosphere, underwater free space QKD is also
faced with two unavoidable problems: one is the attenuation of seawater
channel, and the other is the error of information in communication. The
complex components and special optical properties of seawater become the
biggest challenges for underwater free space QKD research. The light absorption
underwater is an irreversible thermal process whereby photon energy is lost
due to interaction with water molecules or other particulates. It will
reduce the number of received photons, and then influence the secret key generation
rate of QKD. On the other hand, during the propagation, part of the photons
will be scattered and some of them can be received by the receiver.
According to Mie theory, the scattering will change the polarized state of
photons which constitute the qubits, and increase the error of information.
Therefore, the study of light propagation through water channel is crucial
to realize the underwater free space QKD.

Some analyses of underwater optical wireless communication have been
proposed recently \cite{MLanzagorta, RCSmith, LMullen, SJaruwatanadilok}, but their analyses can not be fully applicable to single photon underwater free space QKD. In this paper,
we investigate the optical property of seawater channel and present an analysis
of the feasibility of underwater free space QKD using single photons (based
on polarization coding). We mainly focus on the absorption and scattering of
the seawater channel, and we employ the vector radiative transfer (VRT)
theory which can capture both the attenuation and multiple scattering
effects. To deal with such a random problem, we use Monte Carlo
simulation method to analyze the attenuation of photons, the fidelity of the scattered photons, the quantum bit
error rate and the sifted key rate of underwater quantum
communication, so as to
discuss the feasibility of underwater free space QKD.

\section{Fundamental model and Monte Carlo method}
Seawater is a mixture with extremely complex components. For convenience, the oceanic waters have been divided in Jerlov water types that approximately share the same optical properties \cite{MLanzagorta}. In general, the components only
creating absorption are seawater and colored dissolved organic matter
(CDOM), while the components creating both absorption and scattering are (1)
planktonic components, (2) detrital components, and (3) mineral components \cite{SJaruwatanadilok}. 
The total effects of absorption and scattering are described by the beam
extinction coefficient which is wavelength dependent, $\mu _{e}(\lambda
)=\mu _{a}(\lambda )+\mu _{s}(\lambda )$, where $\mu _{a}(\lambda )$ is the
absorption coefficient and $\mu _{s}(\lambda )$ is the scattering
coefficient, all in units of inverse meters. The photons that reach a
certain location without being scattered or absorbed can be calculated by 
\begin{equation} \label{equ1}
N(l)=N_{0}e^{-\mu _{e}(\lambda )l},
\end{equation}%
where $N_{0}$ is the initial number of photons, and $N(l)$ is the number of
photons after propagating a distance $l$ in water \cite{MLanzagorta}.

\begin{figure}[htbp]
\centering 
\includegraphics[width=6.5cm]{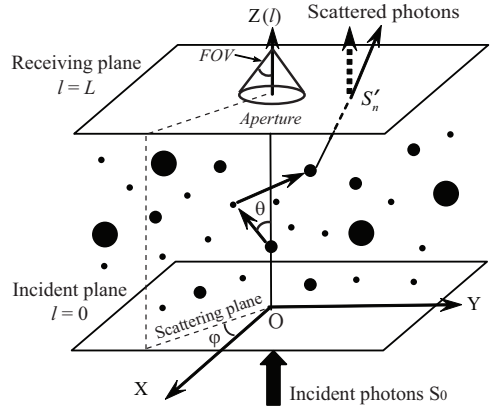}
\caption{The propagation of the polarized photons in the seawater channel, where $\protect\theta$ is the scattering angle, and $\protect\varphi$ is the 
azimuth angle between scattering plane and reference plane. In our model, the receiver with aperture (A) and angle of field of view (FOV) is located in the positive direction of the incident light.}\label{model} 
\end{figure}

We employ the VRT theory which explains the behavior of wave propagation and
scattering in a discrete random medium, and we simulate the optical
propagation in the underwater environment with the Monte Carlo method \cite%
{JCRamella-Roman}. Fig. \ref{model} is the sketch map of the polarized photons
propagating in the seawater channel. In our model, the receiver with aperture (A) and angle of field of view (FOV) is located in the positive direction of the incident light. The polarized state of a photon can be
described by Stokes vector: 
\begin{equation}
S=\left( 
\begin{array}{c}
I \\ 
Q \\ 
U \\ 
V%
\end{array}%
\right) =\left( 
\begin{array}{c}
\langle E_{\parallel }E_{\parallel }^{\ast }+E_{\perp }E_{\perp }^{\ast
}\rangle  \\ 
\langle E_{\parallel }E_{\parallel }^{\ast }-E_{\perp }E_{\perp }^{\ast
}\rangle  \\ 
\langle E_{\parallel }E_{\perp }^{\ast }+E_{\perp }E_{\parallel }^{\ast
}\rangle  \\ 
i\langle E_{\parallel }E_{\perp }^{\ast }-E_{\perp }E_{\parallel }^{\ast
}\rangle 
\end{array}%
\right) ,  \label{equa}
\end{equation}%
where $E_{\parallel }$ and $E_{\perp }$are horizontally and vertically
polarized electric fields, and the symbol $\ast $ and $<>$ denote complex
conjugate and time average \cite{AIshimaru}. In BB84 protocol, Alice and Bob randomly use
four quantum states that constitute two bases, for example, the states $%
|H\rangle $, $|V\rangle $, $|P\rangle $, and $|M\rangle $ are identified as
the linear polarized photons \textquotedblleft horizontal\textquotedblright
, \textquotedblleft vertical\textquotedblright , \textquotedblleft 45%
\textdegree\textquotedblright , and \textquotedblleft 135\textdegree%
\textquotedblright\ respectively. The Stokes vectors of the four quantum states
can be written as $S_{H}=(1,1,0,0)^{T}$, $S_{V}=(1,-1,0,0)^{T}$, $%
S_{P}=(1,0,1,0)^{T}$, and $S_{M}=(1,0,-1,0)^{T}$, where $\mathit{T}$ is the
transpose operator.

We assume the initial position of photons is the origin $O$ 
of the coordinate, and initial direction is z-axis $(0,0,1)$. The
process of single scattering for photons can be described by a
scattering Mueller matrix $M(\theta )$, with $4\times 4$ elements. After the
single scattering, the new Stokes vector changes to $S^{\prime }=M(\theta
)S_{0}$, where $S_0$ is the initial Stokes vector (\textit{l} = 0), and $\theta $ is the scattering angle. For simplicity, the
scattering particles are considered as homogeneous and spherical particles,
the Mueller matrix will have only four independent elements: 
\begin{equation}
M(\theta )=\left( 
\begin{array}{cccc}
m_{1}(\theta ) & m_{2}(\theta ) & 0 & 0 \\ 
m_{2}(\theta ) & m_{1}(\theta ) & 0 & 0 \\ 
0 & 0 & m_{3}(\theta ) & m_{4}(\theta ) \\ 
0 & 0 & -m_{4}(\theta ) & m_{3}(\theta )%
\end{array}%
\right) ,  \label{equb}
\end{equation}%
where the elements $m_{1}(\theta )$, $m_{2}(\theta )$, $m_{3}(\theta )$ and $%
m_{4}(\theta )$ are related to the scattering amplitudes which can be
calculated by Mie theory \cite{JCRamella-Roman}.

To calculate the Mueller matrix, the indexes of refraction of seawater components and
the particle size distributions (PSD) of scattering particles are required.
The indexes of refraction can be found in \cite{SJaruwatanadilok} and will not be
described here. Usually a single distribution function $N(D)$ can be used to
describe the PSD, which complies with the Jungle (also known as hyperbolic) cumulative size
distribution \cite{HBader}: 
\begin{equation}
N(D) = K(\frac{D}{D_0})^{-\epsilon},
\end{equation}
where $D_0$ is a reference diameter for which the number concentration is $K$. $\epsilon$ is different for different types of particles, and usually
ranges from 3 to 5 typically. In order to determine the diameters of
scattering particles, we use Monte Carlo method to choose a random diameter $%
D$ according to the distribution $N(D)$.

The distance between two successive scatterings of a photon is usually called
free path \cite{JCRamella-Roman}, which is determined by 
\begin{equation}
\Delta L = - \frac{ln(\eta)}{\mu_e},
\end{equation}
where $\eta\in(0, 1]$ is a random number.

For multiple scattering, the reference plane is different between
two scatterings. For convenience, we can take scattering plane as the
reference plane. The Stokes vectors of photons are rotated by an azimuth angle $\varphi$ at each scattering step. In the case
of $n$ times scatterings, we obtain the last Stokes vector 
\begin{equation}  \label{equc}
S_n^{\prime }= M(\theta_n) R(\varphi_n) \cdots M(\theta_1) R(\varphi_1) S_0,
\end{equation}
where $R(\varphi)$ is a
rotation matrix \cite{JCRamella-Roman}: 
\begin{equation}  \label{equd}
R(\varphi) = \left( 
\begin{array}{cccc}
1 & 0 & 0 & 0 \\ 
0 & cos(2\varphi) & sin(2\varphi) & 0 \\ 
0 & - sin(2\varphi) & cos(2\varphi) & 0 \\ 
0 & 0 & 0 & 1%
\end{array}
\right).
\end{equation}

\begin{figure}[htb]
\subfigure[The scattering angle (range from 0\textdegree to 180\textdegree).]{ 
    \begin{minipage}[b]{0.5\textwidth} 
      \centering 
      \includegraphics[width=6.5cm]{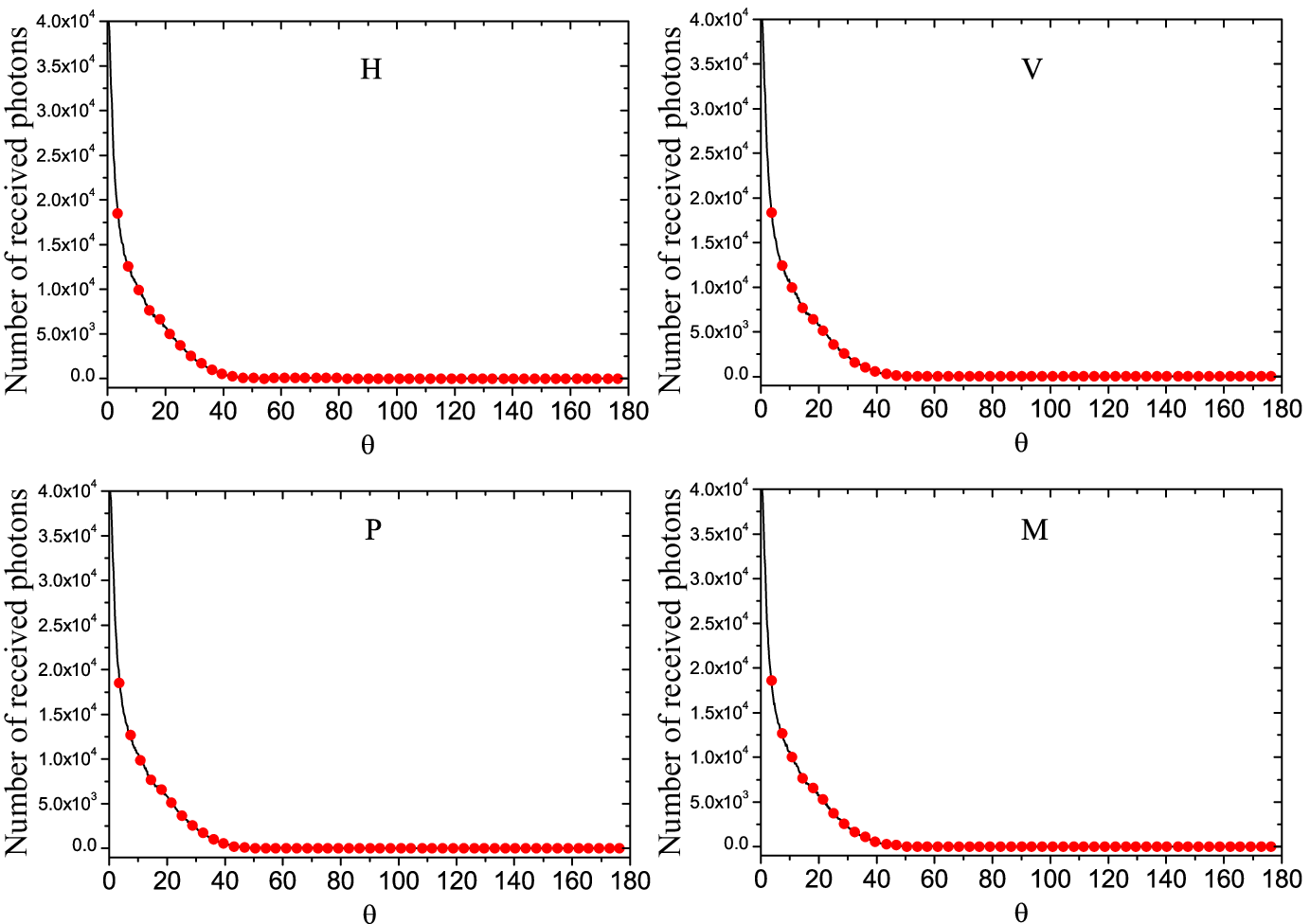} 
    \end{minipage}} 
\subfigure[The azimuth angle (range from 0\textdegree to 360\textdegree).]{ 
    \begin{minipage}[b]{0.5\textwidth} 
      \centering 
      \includegraphics[width=6.5cm]{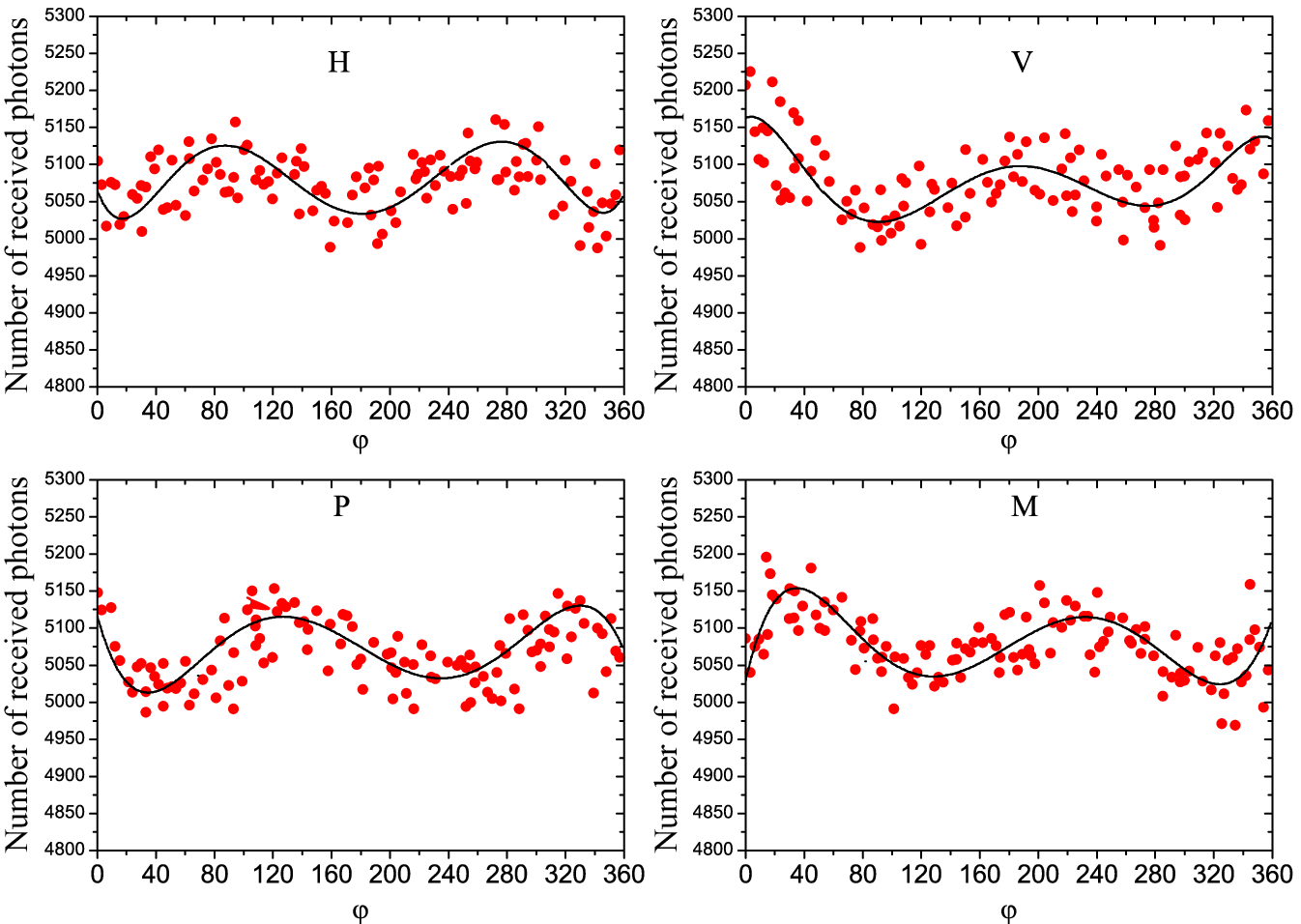} 
    \end{minipage}}
\caption{(Color online) The distribution of the scattering angle and the
azimuth angle. }\label{scazai} 
\end{figure}

According to the above equations (\ref{equa}, \ref{equb}, \ref{equc}, and \ref%
{equd}), we can obtain the scattering phase function: 
\begin{equation}
P(\theta,\varphi) = m_1(\theta) I +m_2(\theta)[ cos(2\varphi) Q +
sin(2\varphi) U ],
\end{equation}
which can be used to sample the scattering angle $\theta$
and the azimuth angle $\varphi$ by the rejection method \cite%
{JCRamella-Roman}. Fig. \ref{scazai} shows the distribution of the scattering angle $%
\theta$ and the azimuth angle $\varphi$ for four polarized states, where red dots denote the statistical results obtained with Monte
Carlo method, and black lines are the fitting curve related to the
statistical results. The total number of photons we use is $10^6$. 

Other two important steps in our simulation are the determination of lifetime of photons and the boundary problems. We define the
weight of lifetime of photons by 
\begin{equation}
W(\Delta L)=\frac{\mu _{s}^{\prime }}{\mu _{s}^{\prime }+\mu _{a}^{\prime }}%
e^{-\mu _{m}\Delta L},
\end{equation}%
where $\mu _{s}^{\prime }$ and $\mu _{a}^{\prime }$ are the scattering
coefficient and the absorption coefficient of the scattering particles, $\mu
_{m}$ is the absorption coefficient of the seawater and dissolved substances
in it. $W$ can be considered as the probability of the surviving photons
after moving a propagation distance $\Delta L$. In order to determine the
lifetime of photons, a uniform random number $\xi $ between 0 and 1 should
be generated firstly. The photon is regarded as being absorbed if $\xi >W$,
then the next photon will be launched. Otherwise, the photon is regarded as
being scattered if $\xi <W$, then it will continue to propagate. The other
problem is when and where the photon arrives at the boundary (receiving
plane, \textit{l} = L). In our simulation, the positions of a photon are recorded constantly, and the photon arrives at the boundary when $\mathit{l}%
\geq L$. Photons can be received or not by the receiver depending on the
aperture and FOV size of the receiver.

\section{Simulation results and discussions}

The bit rate and the bit error rate are two crucial points for any
communication process. They depend on both the performance of the
communication system and the properties of communication channel. The communication system is not the focus of this article. We mainly discuss the influence of
the seawater channel. In underwater free space QKD, the attenuation of
the seawater channel will reduce the bit rate, while scattered photons and the
noise in the channel will increase the bit error rate. One cannot
increase the signal power in order to have a good enough signal to noise
ratio (SNR) since the signal transmitted by seawater in QKD is ideally single photons (or weak coherent pulses with very low mean photon number in many
realistic implementations). Therefore the available methods to implement
underwater free space QKD are reducing the water channel attenuation,
scattered photons, and the background noise.

In our Monte Carlo simulations, we mainly investigate the propagation of photons in underwater free space QKD with the limitation of receiving aperture diameter($10 cm$ - $50 cm$) and FOV ($10 \degree$ - $30 \degree$)\cite{JWGiles}. We mainly choose the size of Mie scattering particles from $1\mu m$ to $200\mu m$ considering that plankton is widespread in Jerlov type I - III water \cite{CDMobley}, and we ignore CDOM and other components in ocean water. To simplify the calculation we take the mean complex refractive index of particles as $1.41 - 0.00672i$\cite{CDMobley} (in the case of plankton).

\subsection{Light attenuation}
\begin{figure}[htbp]
\centering 
\includegraphics[width=6.5cm]{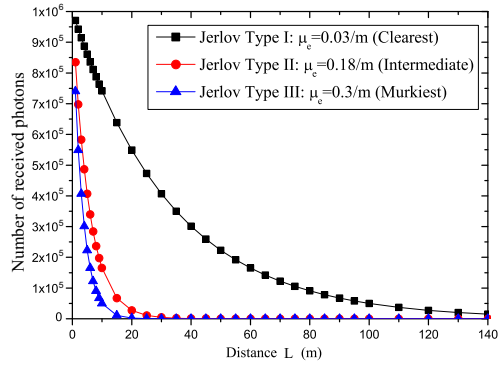}
\caption{(Color online) Optical attenuation for $\protect\lambda = 480 nm $
in clearest (Jerlov type I, $\mu_e=0.03/m$), intermediate (Jerlov type II, $\mu_e=0.18/m$), and murky (Jerlov type III, $\mu_e=0.3/m$) ocean waters.}\label{attenuation} 
\end{figure}

In order to reduce the seawater channel attenuation, free space QKD should work
in the blue-green light wavelengths because it suffers less attenuation in
water compared to other colors. In our Monte Carlo simulations, we take the wavelength of photons as $480 nm$, and we investigate the effects of the total extinction coefficient on the received photons, where the absorption coefficient of pure seawater is $0.0176/m$ \cite{CDMobley} at $\lambda = 480 nm$ and the absorption coefficient and the scattering coefficient of plankton can be
calculated by Mie theory. Fig. \ref{attenuation} shows that the number of received photons changes at different communication distances in three types of ocean waters, where the solid lines are the
theoretical values, eq. (\ref{equ1}), while the scatter dots are results of our simulation in
receiving aperture of $10cm$ and FOV of $10\degree$($175 mrad$). They are almost coincide
with one another in the same water type. 

As expected, the number of received photons
decreases with the increase of distance, therefore the secret key generation
rate (related to the bit rate) of QKD decreases. It can be observed that,
for example, in the clearest oceanic water, only $5\%$ of photons reach
a distance of $100 m$. The situations of the intermediate and murky ocean water are even worse, for example, only $0.01\%$ of photons can reach a distance of only $50 m$ in the intermediate ocean water and $30 m$ in the murky ocean water. Therefore underwater free space QKD is suitable only for short-range underwater communication, especially in the intermediate and murky ocean water the communication distance is greatly restricted. 

\begin{figure}[tbh]
\subfigure[$FOV = 10\degree (175 mrad)$.]{ 
    \begin{minipage}[b]{0.5\textwidth} 
      \centering 
      \includegraphics[width=6.5cm]{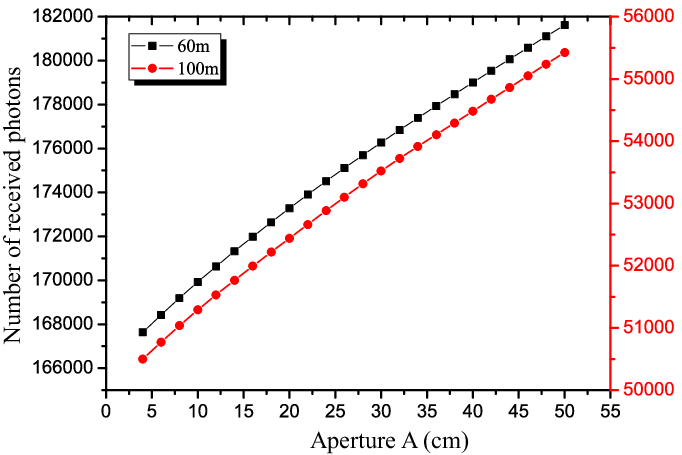} 
    \end{minipage}} 
\subfigure[$A = 20 cm$.]{ 
    \begin{minipage}[b]{0.5\textwidth} 
      \centering 
      \includegraphics[width=6.5cm]{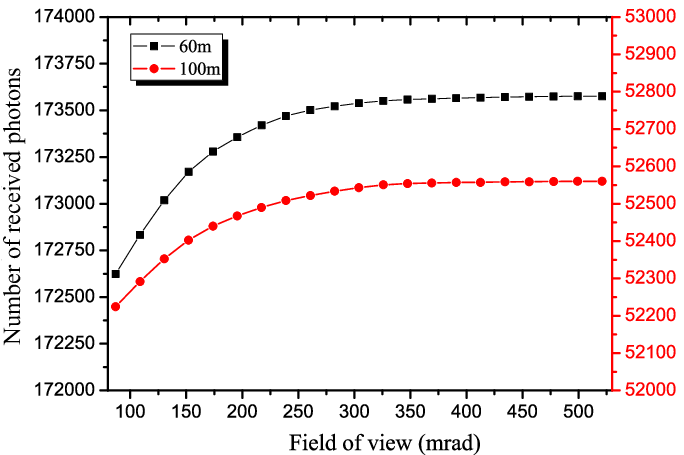} 
    \end{minipage}}
\caption{(Color online) Number of received photons as functions of the receiving aperture and FOV
in Jerlov Type I ocean water, where black square denotes the communication distance is $60m$, red circle dot denotes the communication distance is $100m$.}\label{number} 
\end{figure}

Fig. \ref{number} shows that the number of received photons increases with the increases of aperture and FOV, but the increases are not very great. Especially when the FOV of receiver is more than $250 mrad$, the number of received photons is almost unchanged. Indeed, the increased photons are all scattered photons, therefore the receiver with bigger aperture and bigger FOV will receive more scattered photons. 

\subsection{Fidelity of the received scattered photons}
In quantum communication, fidelity is usually used to describe how similar
two quantum states are. Suppose an arbitrary quantum system is in one of states $|\psi\rangle_i$ with respective
probabilities $p_i$. The density operator for the system is defined as $\rho
= \sum p_i |\psi\rangle_i \langle\psi|$. The fidelity between a pure state $%
|\psi\rangle $ and the state is \cite{MANielsen}
\begin{equation}  \label{equf}
F(|\psi\rangle, \rho) = \sqrt{\langle\psi| \rho |\psi\rangle}.
\end{equation}

\begin{figure}[tbh]
\subfigure[$FOV = 10\degree (175 mrad)$.]{ 
    \begin{minipage}[b]{0.5\textwidth} 
      \centering 
      \includegraphics[width=6.5cm]{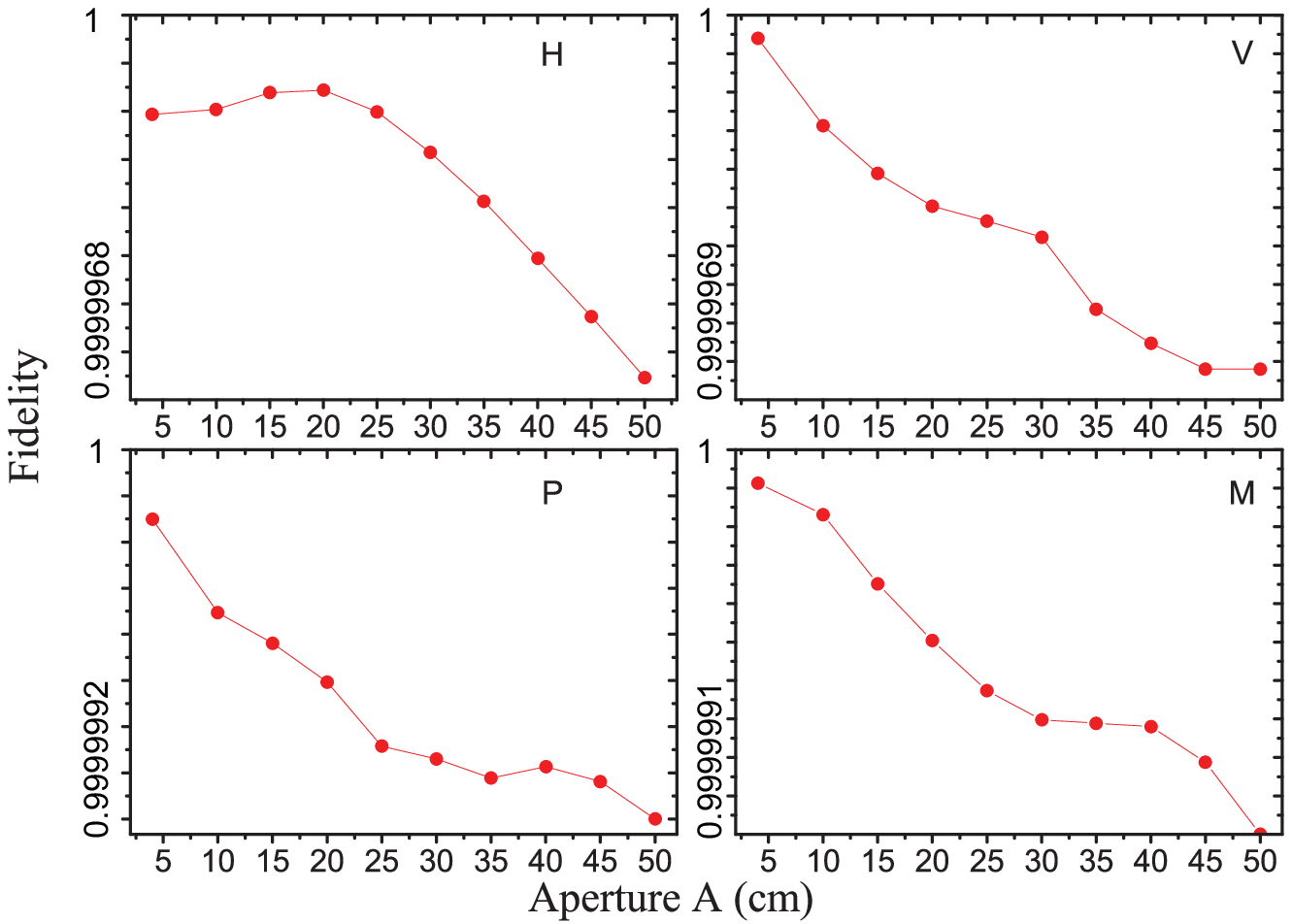} 
    \end{minipage}} 
\subfigure[$A = 20 cm$.]{ 
    \begin{minipage}[b]{0.5\textwidth} 
      \centering 
      \includegraphics[width=6.5cm]{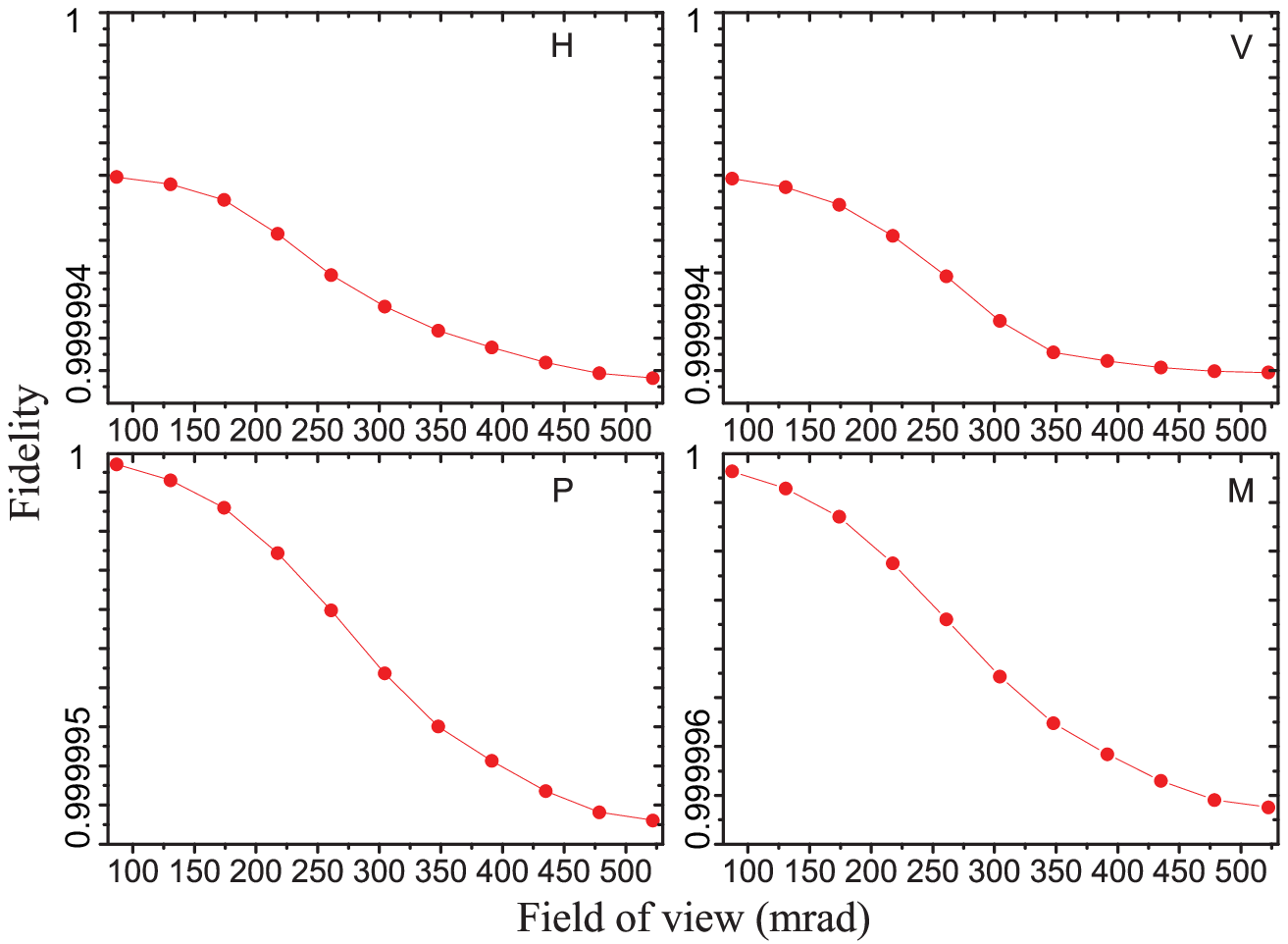} 
    \end{minipage}}
\caption{(Color online) The fidelity of the received scattered photons as functions of the receiving aperture and FOV in Jerlov Type I ocean water, where the communication distance is 60m, and H, V, P, M denote the linear polarized states of \textquotedblleft 0\textdegree\textquotedblright
, \textquotedblleft 90\textdegree\textquotedblright , \textquotedblleft 45\textdegree\textquotedblright , and \textquotedblleft 135\textdegree\textquotedblright\ respectively.}\label{fidelity1} 
\end{figure}
Fig. \ref{fidelity1} shows the fidelity of the received scattered photons as functions of receiving aperture and FOV in the clearest ocean water. With the increase of aperture and FOV respectively, the fidelities of the four linear polarized states decrease slowly.

\begin{figure}[htbp]
\centering 
\includegraphics[width=9cm]{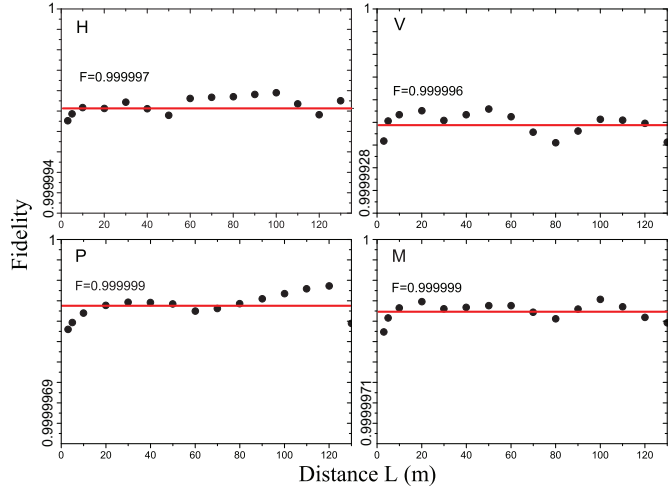}
\caption{(Color online) The fidelity of the received scattered photons as a function of communication distance in
Jerlov Type I ocean water, where $A = 10 cm$, $FOV = 10 \degree (175 mrad)$.}\label{fidelity2} 
\end{figure}
We also calculate the fidelity of the received scattered photons at different communication
distances in the clearest
ocean water (shown in Fig. \ref{fidelity2}). The results show that the
polarized states of the received scattered photons keep almost invariant (almost equal to 1) with the increase of distance. The reason why the fidelity is close to 1 is that, with
the limitation of the receiving aperture and FOV, most of the received photons are
not scattered. Even if the scattered photons are received, the polarized states of these received photons are not changed significantly due to the extremely small scattering angles of these received scattered photons. 

\subsection{Noise and quantum bit error rate}
The quantum bit error rate (QBER) is an important parameter used to characterize the QKD system \cite{NGisin, VScarani09}. It describes the
probability of false detection in total probability of detection per pulse.
For underwater quantum communication, the false detections result from 
background noises, scattered photons and imperfection of optical devices, etc. In this paper, only scattered photons and background light noise under water are considered. For a typical BB84 QKD system, if we ignore the dark current and the efficiency of the detector, the QBER is given by \cite{MLanzagorta, DJRogers}

\begin{equation}
QBER = \frac{Error}{N+2Error},
\end{equation}
where $N$ is the number of received signal photons which can be calculated in our simulation, and
\begin{equation}
Error = Scatter_{Error} + \frac{\pi^2 R_d A^2 \Delta t' \lambda [1-cos(FOV)]}{8 h c \Delta t},
\end{equation}
where $Scatter_{Error}$ is the error number of received scattered photons which can be calculated in our simulation too, $R_d$ is the irradiance of the environment, $\Delta t'$ is the receiver gate time, $h$ is Planck's constant, $c$ is the speed of light in vacuum, $\Delta t$ is the bit period. In our simulations we choose typical values of currently available free space BB84 QKD systems \cite{MLanzagorta, DJRogers}: $A=10 cm$, $FOV=10\degree$, $\lambda=480 nm$, $\Delta t'=200 ps$, $\Delta t=35 ns$. 

There are two kinds of background light
in seawater channel. One is the radiation or reflection from the sun, moon and stars.
The light can be scattered by molecules
and particles in water when it travels through the sea, then collected by the receiver as background noises. Another kind of background light is created by the illuminant in the water,
such as marine luminous organisms. The receiver should be kept away from these
luminous bodies in underwater free space QKD. We are only interested in the background light from the sun, moon and stars. Table \ref{Tirradiances} gives some typical values of total irradiances reaching sea level for various environmental conditions \cite{CDMobley}. 
\begin{table}[h]
  \caption{Typical total irradiances at sea level in the visible wavelength band (400-700nm) \label{Tirradiances}}
  \begin{center}
    \begin{tabular}{lc}
    \hline
    Environment & Irradiance ($W/m^2$)\\
    \hline
    clear atmosphere, full moon near the zenith & $1 \times 10^{-3}$\\
    clear atmosphere, starlight only & $1 \times 10^{-6}$\\
    cloudy night & $1 \times 10^{-7}$\\
    \hline
    \end{tabular}
  \end{center}
\end{table}

Under typical oceanic conditions, for which the incident light is
provided by the sun, moon and stars, the various irradiances decreases
approximately exponentially with depth, at least when far enough below the
surface (and far enough above the bottom, in shallow water) to be free of
boundary effects \cite{CDMobley}. It is convenient to write the depth dependence of $R_d(\lambda, z)$ as 
\begin{equation}
R_d(\lambda, z) = R_d(\lambda, 0) e^{- \mu_d (\lambda, z) z},
\end{equation}
where $\mu_d (\lambda, z)$ is the average diffuse attenuation coefficient over the depth interval from $0$ to $z$. We take $\mu_d \approx 0.019 /m$ ($\lambda = 480 nm$) in the Jerlov type I water \cite{CDMobley}. In our simulations, we assume that the receiver is placed toward the sea level, and the laser device is above the receiver.

\begin{figure}[tbh]
\subfigure[$L = 60 m$, $FOV = 10 \degree (175 mrad)$.]{ 
    \begin{minipage}[b]{0.5\textwidth} 
      \centering 
      \includegraphics[width=6.5cm]{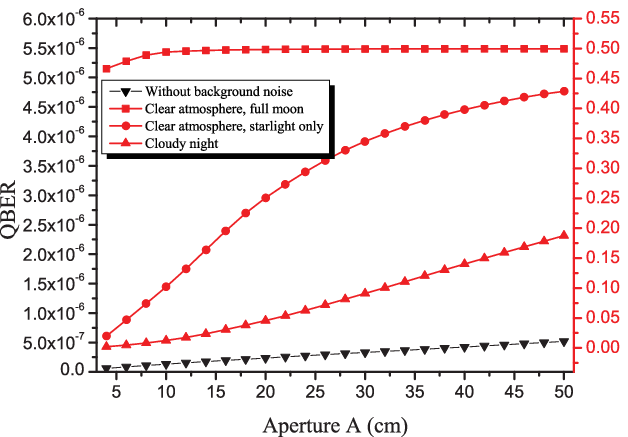} 
    \end{minipage}} 
\subfigure[$L = 60 m$, $A = 20 cm$.]{ 
    \begin{minipage}[b]{0.5\textwidth} 
      \centering 
      \includegraphics[width=6.5cm]{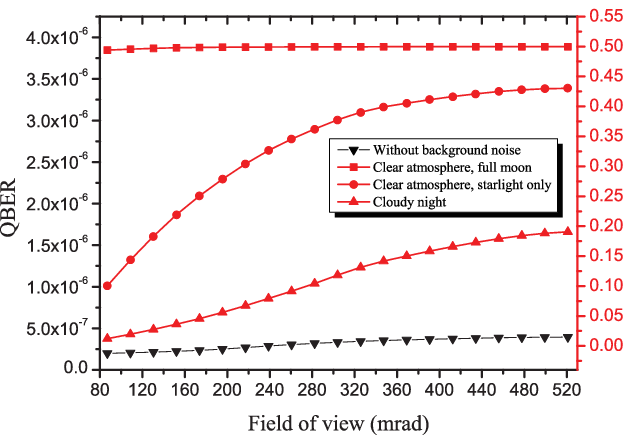} 
    \end{minipage}}
\caption{(Color online) QBER as functions of the receiving aperture and FOV
in Jerlov Type I ocean water, where the black reverse triangle denotes QBER without
background noise, the red square, circle dot, triangle respectively denote QBER in different environmental conditions of full moon, starlight only, and cloudy night.}\label{QBER1}
\end{figure}

\begin{figure}[htbp]
\centering 
\includegraphics[width=7cm]{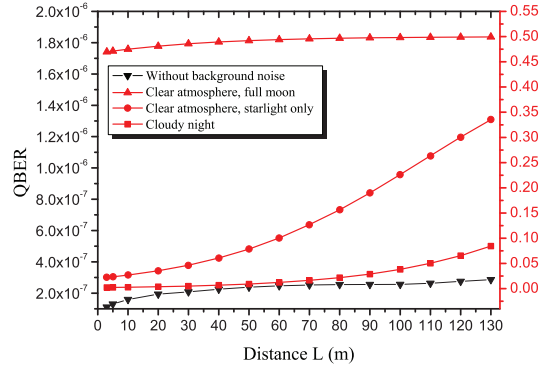}
\caption{(Color online) QBER as a function of communication distance in Jerlov Type I ocean water, where $A = 10 cm$, $FOV = 10 \degree (175 mrad)$, and the black reverse triangle denotes QBER without
background noise, the red triangle, circle dot, square respectively denote QBER in different environmental conditions of full moon, starlight only, and cloudy night.}\label{QBER2}
\end{figure}
We investigate the QBER as functions of the receiving aperture and FOV at the same communication distance in four different environmental conditions (shown in Fig. \ref{QBER1}), and we assume that the receiver is in $200 m$ deph under sea level. With the increase of the aperture and FOV, both the scattered photons and background noises will increase, however, the effects of scattered photons are far less than that of background light noises. Fig. \ref{QBER1} shows that QBER is close to 0 without considering the background noise, and QBER increases significantly with the increase of the receiving aperture and FOV in other three environmental conditions. Hence, small aperture and FOV are suitable for underwater quantum communication in order to suppress QBER. On the other hand, the numbers of total received photons reduce with the increase of
communication distance, that is to say, the signal photons received by the receiver decrease. Thus the SNR of quantum communication decreases due to the
increasing proportion of the background light. Its influence is especially
obvious with the increasing distance. Fig. \ref{QBER2} shows the results of QBER with
the increase of communication distance. 

\begin{figure}[htb]
\centering 
\includegraphics[width=7cm]{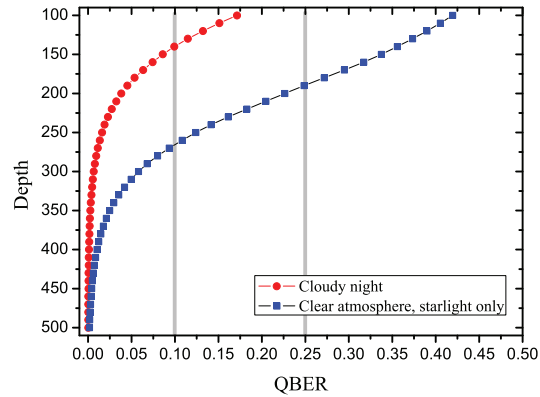}
\caption{(Color online) Changes of the QBER with different depths of seawater in 
Jerlov Type I ocean water, where $A = 10 cm$, $FOV = 10 \degree (175 mrad)$, $\protect\mu_e
= 0.03/m$, and the communication distance of the underwater free space QKD is about $100 m$. The circle dot, square respectively denote QBER in different environmental conditions of cloudy night and starlight only.}\label{QBER3}
\end{figure}
In particular, for the case of BB84 protocol, the system is secure against a
sophisticated quantum attack if $QBER\leq 10\%$, and the system is secure against a simple intercept-resend attack if $QBER\leq 25\%$ \cite{MLanzagorta}. From the above discussions, it can be observed that in the environmental condition of starlight only most secure single photon underwater BB84 QKD is feasible up to about $60 m$ in the clearest ocean water (the receiver is in $200 m$ depth), while BB84 QKD secure against simple intercept-resend attacks is feasible up to about $107 m$ in the clearest ocean waters. In addition, in the environmental condition of cloudy night most secure single photon underwater BB84 QKD is feasible up to more than $130 m$ in the clearest ocean water (the receiver is in $200 m$ depth).

Considering that the noise of background light is different in different depths of seawater, we choose a range of $100 m$ - $500 m$ seawater depths to estimate the QBER in two different environmental conditions. Fig. \ref{QBER3} shows the changes of the QBER with different depths of seawater, where $A = 10 cm$, $FOV = 10 \degree (175 mrad)$, and the communication distance of the underwater free space QKD is about $100 m$.

It can be observed that the QBER in the two environmental conditions reduces with the increase of
seawater depth, and in the same depth the QBER at cloudy night is less than that in the condition of starlight only. In order to implement $100 m$ most secure single photon underwater BB84 QKD in Jerlov Type I ocean water, the receiver should be respectively sunk to more than $130$ meters and $260$ meters under sea level in the environmental condition of cloudy night and starlight only. In addition, in the environmental condition of starlight only the receiver can be sunk to more than $185$ meters under sea level to implement $100 m$ QKD with the secure against simple intercept-resend attacks.

\subsection{The sifted key generation rate}

\begin{figure}[htb]
\centering 
\includegraphics[width=7cm]{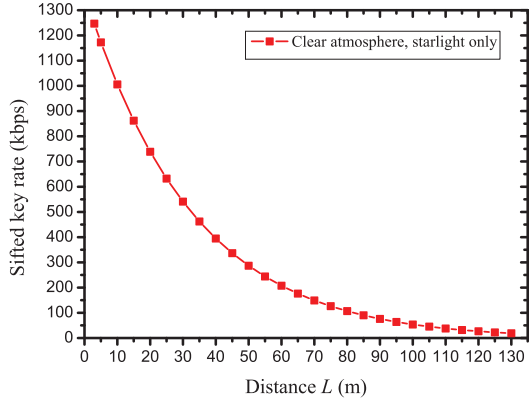}
\caption{(Color online) The sifted key generation rate in the environmental condition of starlight only as a function of communication distance in Jerlov Type I ocean water. Where $A = 10 cm$, $FOV = 10 \degree (175 mrad)$, $\protect\mu_e = 0.03/m$, $f = 1/35 GHz$, $\langle N\rangle = 0.1$, $depth = 200 m$.}\label{siftedkey}
\end{figure}
In practical QKD system, take BB84 protocol for example, after attenuation
and sifting, the sifted key generation rate is given by \cite{NGisin, VScarani09}
\begin{equation}
\kappa = f \cdotp \langle N \rangle \cdotp (1 - a) \cdotp (1 - QBER) \cdotp %
DE / 2,
\end{equation}
where $f$ is laser pulse frequency, $\langle N \rangle$ is mean photons
number per pulse, $a$ is attenuation rate of channel, $QBER$ is bit error
rate, and $DE$ is detection efficiency. Take the condition of starlight only (shown in fig. \ref{QBER2}) for example, we can estimate the sifted key generation rate of underwater free space QKD. If we ignore the detection efficiency, Fig. \ref{siftedkey} shows the sifted key generation rate as a function of communication distance in Jerlov Type I ocean water. From the above simulation results, we estimate the sifted key generation rate $\kappa \approx 207 kbps$ at the communication distance of $60 m$, and $\kappa \approx 45 kbps$ at the communication distance of $107 m$ in the clearest ocean water (the receiver is in $200 m$ depth) in the environmental condition of starlight only. The keys generated in this situation could be used for encrypting most audio information and some low bit rate
video information in underwater communications.

\section{Conclusion}
In summary, we present an analysis of the optical propagation for underwater
free space QKD. With the vector radiative transfer theory and Monte Carlo
simulations, we obtain the attenuation of photons, the fidelity of the scattered photons, the QBER and the sifted key generation rate of underwater quantum
communication, and discuss the
feasibility of underwater free space QKD. The results show that the attenuation and background noise are the important factors of influencing underwater
free space QKD. It can be observed that most secure single photon underwater free space QKD is feasible
in the clearest ocean water.

In practical underwater free space QKD, reducing the attenuation and
background noise is the crucial technology. Light wavelength should be
chosen in blue-green wavelength region in order to reduce the attenuation of seawater channel. Background noise can be reduced by implementing a
strong filtering in the spatial, spectral and temporal domains \cite{CBonato, ELMiao}. In principle, the receiving aperture and FOV
can suppress the background noise effectively, so we can control the size
of aperture and FOV to implement a spatial filter. Usually a narrow band filter must be used
before the receiver to prevent the background
light. Another effective method to decrease the
number of background photons is using a time gate. Only when the
signal photons are expected to arrive, is a narrow time gate opened to allow
them to enter the receiver, thus noise photons arriving outside the time
window are blocked.

\section*{Acknowledgments}

This work was supported by the National Natural Science Foundation of China
(Grant No. 60677044, 11005099) and the Fundamental Research Funds for the
central universities (Grant No. 201313012).

\end{document}